\documentclass[10pt,preprint]{aastex}

\newcommand{\latin}[1]{{#1}}

\newcommand{\eg}{\latin{e.g.}}

\def\simless{\mathbin{\lower 3pt\hbox
	{$\,\rlap{\raise 5pt\hbox{$\char'074$}}\mathchar"7218\,$}}} 
\def\simgreat{\mathbin{\lower 3pt\hbox
	{$\,\rlap{\raise 5pt\hbox{$\char'076$}}\mathchar"7218\,$}}} 

\newcounter{thefigs}
\newcommand{\fignum}{\arabic{thefigs}}

\newcounter{thetabs}

\newcounter{address}

\slugcomment{Submitted to \apj}

\shortauthors{Blanton {\it et al.} (2004)}
\shorttitle{Scale dependence of clustering}

\begin{document}

\title{The scale-dependence of relative galaxy bias: \\
encouragement for the ``halo model'' description\altaffilmark{\ref{SDSS}}}

\author{
Michael R. Blanton\altaffilmark{\ref{NYU}}, 
Daniel Eisenstein\altaffilmark{\ref{Steward}},
David W. Hogg\altaffilmark{\ref{NYU}}, and 
Idit Zehavi\altaffilmark{\ref{Steward}}
}

\altaffiltext{1}{Based on observations obtained with the
Sloan Digital Sky Survey\label{SDSS}} 
\setcounter{address}{2}
\altaffiltext{\theaddress}{
\stepcounter{address}
Center for Cosmology and Particle Physics, Department of Physics, New
York University, 4 Washington Place, New
York, NY 10003
\label{NYU}}
\altaffiltext{\theaddress}{
\stepcounter{address}
Steward Observatory, 933 N. Cherry Ave., Tucson, AZ 85721
\label{Steward}}
\altaffiltext{\theaddress}{
\stepcounter{address}
Princeton University Observatory, Princeton,
NJ 08544
\label{Princeton}}
\altaffiltext{\theaddress}{
\stepcounter{address}
Apache Point Observatory,
2001 Apache Point Road,
P.O. Box 59, Sunspot, NM 88349-0059
\label{APO}}

\begin{abstract}
We investigate the relationship between the colors,
luminosities, and environments of galaxies in the Sloan
Digital Sky Survey spectroscopic sample, using environmental
measurements on scales ranging from $0.2$ to $6$ $h^{-1}$
Mpc. We find: (1) that the relationship between color and
environment persists even to the lowest luminosities we
probe ($M_r - 5 \log_{10} h \sim -14$); (2) at luminosities
and colors for which the galaxy correlation function has a
large amplitude, it also has a steep slope; and (3) in
regions of a given overdensity on small scales (1 $h^{-1}$
Mpc), the overdensity on large scales (6 $h^{-1}$ Mpc) does
not appear to relate to the recent star formation history of
the galaxies. Of these results, the last has the most
immediate application to galaxy formation theory. In
particular, it lends support to the notion that a galaxy's
properties are related only to the mass of its host dark
matter halo, and not to the larger scale environment.
\end{abstract}


\section{Motivation}
\label{motivation}

Recent investigations of the large scale distribution of galaxies in
the Sloan Digital Sky Survey (SDSS; \citealt{abazajian04a}) have
revealed a complex relationship between the properties of galaxies,
(such as color, luminosity, surface brightness, and concentration) and
their environments (\citealt{hogg03b, blanton03q, berlind04a}). These
and other investigations using the SDSS (\citealt{zehavi02a,
zehavi04a, kauffmann04a}) and the Two-degree Field Galaxy Redshift
Survey (\citealt{norberg02a, balogh04a}) have found that galaxy
clustering is a function both of star formation history and of
luminosity. For low luminosity galaxies, clustering is a strong
function of color, while for luminous galaxies clustering is a strong
function of luminosity. For red galaxies, clustering is a
non-monotonic function of luminosity, peaking at both high and low
luminosities. Although galaxy clustering correlates also with surface
brightness and concentration, \citet{blanton03q} and
\citet{kauffmann04a} show that galaxy environment is independent of
these properties at fixed color and luminosity.  Thus, color and
luminosity --- measures of star formation history --- appear to have a
more fundamental relationship with environment than do surface
brightness and concentration --- measures of the distribution of stars
within the galaxy.

Some of the investigations above have explored the scale dependence of
these relationships. Studies of the correlation function, such as
\citet{norberg02a} and \citet{zehavi04a}, can address this question,
but do not address directly whether the density on large scales is
related to galaxy properties {\it independent} of the relationships
with density on small scales.  If only the {\it masses} of the host
halos of galaxies strongly affect their properties, then we expect no
such independent relationship between galaxy properties and the large
scale density field.  Thus, it is important to examine this issue in
order to test the assumptions of the ``halo model'' description of
galaxy formation and of semi-analytic models that depend only on the
properties of the host halo ({\it e.g.},
\citealt{kauffmann93a, seljak00, white01a, berlind02a}).
Recent studies of this question have come to conflicting
conclusions. For example, \citet{balogh04a} have concluded from their
analysis of SDSS and 2dFGRS galaxies that the equivalent width of
H$\alpha$ is a function of environment measured on scales of 1.1
$h^{-1}$ Mpc and 5.5 $h^{-1}$ Mpc independently of each other. On the
other hand, \citet{kauffmann04a} find that at fixed density at scales
of 1 $h^{-1}$ Mpc, the distribution of D4000 (a measure of the age of
the stellar population) is not a strong function of density on larger
scales.

Here we address the dependence on scale of the relative bias of SDSS
galaxies. Section \ref{data} describes our data set. Section
\ref{results} explores how the relationship between the color, luminosity,
and environments of galaxies depends on scale. Section \ref{bluefrac}
resolves the discrepancy noted in the previous paragraph between
\citet{kauffmann04a} and \citet{balogh04a}, finding that only small
scales are important to the recent star formation history of
galaxies. Section \ref{summary} summarizes the results. 

Where necessary, we have assumed cosmological parameters $\Omega_0 =
0.3$, $\Omega_\Lambda = 0.7$, and $H_0 = 100 h$ km s$^{-1}$ Mpc$^{-1}$
with $h=1$.

\section{Data}
\label{data}

\subsection{SDSS}

The SDSS is taking $ugriz$ CCD imaging of $10^4~\mathrm{deg^2}$ of the
Northern Galactic sky, and, from that imaging, selecting $10^6$
targets for spectroscopy, most of them galaxies with
$r<17.77~\mathrm{mag}$ \citep[\eg,][]{gunn98a,york00a,abazajian03a}.
Automated software performs all of the data processing: astrometry
\citep{pier03a}; source identification, deblending and photometry
\citep{lupton01a}; photometricity determination \citep{hogg01a};
calibration \citep{fukugita96a,smith02a}; spectroscopic target
selection \citep{eisenstein01a,strauss02a,richards02a}; spectroscopic
fiber placement \citep{blanton03a}; and spectroscopic data reduction.
An automated pipeline called {\tt idlspec2d} measures the redshifts
and classifies the reduced spectra (Schlegel et al., in preparation).

The spectroscopy has small incompletenesses coming primarily from (1)
galaxies missed because of mechanical spectrograph constraints
\citep[6~percent;][]{blanton03a}, which leads to a slight
under-representation of high-density regions, and (2) spectra in which
the redshift is either incorrect or impossible to determine
($<1$~percent).  In addition, there are some galaxies ($\sim
1$~percent) blotted out by bright Galactic stars, but this
incompleteness should be uncorrelated with galaxy properties.

\subsection{NYU-VAGC}

For the purposes of computing large-scale structure and galaxy
property statistics, we have assembled a subsample of SDSS galaxies
known as the NYU Value Added Galaxy Catalog (NYU-VAGC;
\citealt{blanton04a}). One of the products of that catalog is a low
redshift catalog. Here we use the version of that catalog
corresponding to the SDSS Data Release 2 (DR2).

The low redshift catalog has a number of important features which are
useful in the study of low luminosity galaxies. Most importantly:
\begin{enumerate}
\item We have checked by eye all of the images and spectra of low
luminosity ($M_r-5\log_{10} h > -15$) or low redshift ($z<0.01$)
galaxies in the NYU-VAGC. Most significantly, we have trimmed those
which are ``flecks'' incorrectly deblended out of bright galaxies; for
some of these cases, we have been able to replace the photometric
measurements with the measurements of the parents. For a full
description of our checks, see
\citet{blanton04a}.
\item For galaxies which were shredded in the target version of the
deblending, the spectra are often many arcseconds away from the
nominal centers of the galaxy in the latest version of the photometric
reductions. We have used the new version of the deblending to decide
whether these (otherwise non-matched spectra) should be associated
with the galaxy in the best version.
\item We have estimated the distance to low redshift objects using the
\citet{willick97a} model of the local velocity field (using $\beta =
0.5$), and propagated the uncertainties in distance into uncertainties
in absolute magnitude. 
\end{enumerate}

For the purposes of our analysis below, we have matched this sample to
the results of \citet{tremonti04a}, who measured emission line fluxs
and equivalent widths for all of the SDSS spectra. Below, we use their
results for the H$\alpha$ equivalent width.

The range of distances we include is $10 < d < 150$ $h^{-1}$ Mpc,
making the sample volume limited for galaxies with $M_r-5\log_{10} h <
-18.5$.  The total completeness-weighted effective area of the sample,
excluding areas close to Tycho stars, is 2220.9 square degrees. The
catalog contains 28,089 galaxies. \citet{blanton04b} have investigated
the luminosity function, surface brightness selection effects, and
galaxy properties in this sample.

\begin{figure}
\figurenum{\fignum}
\plotone{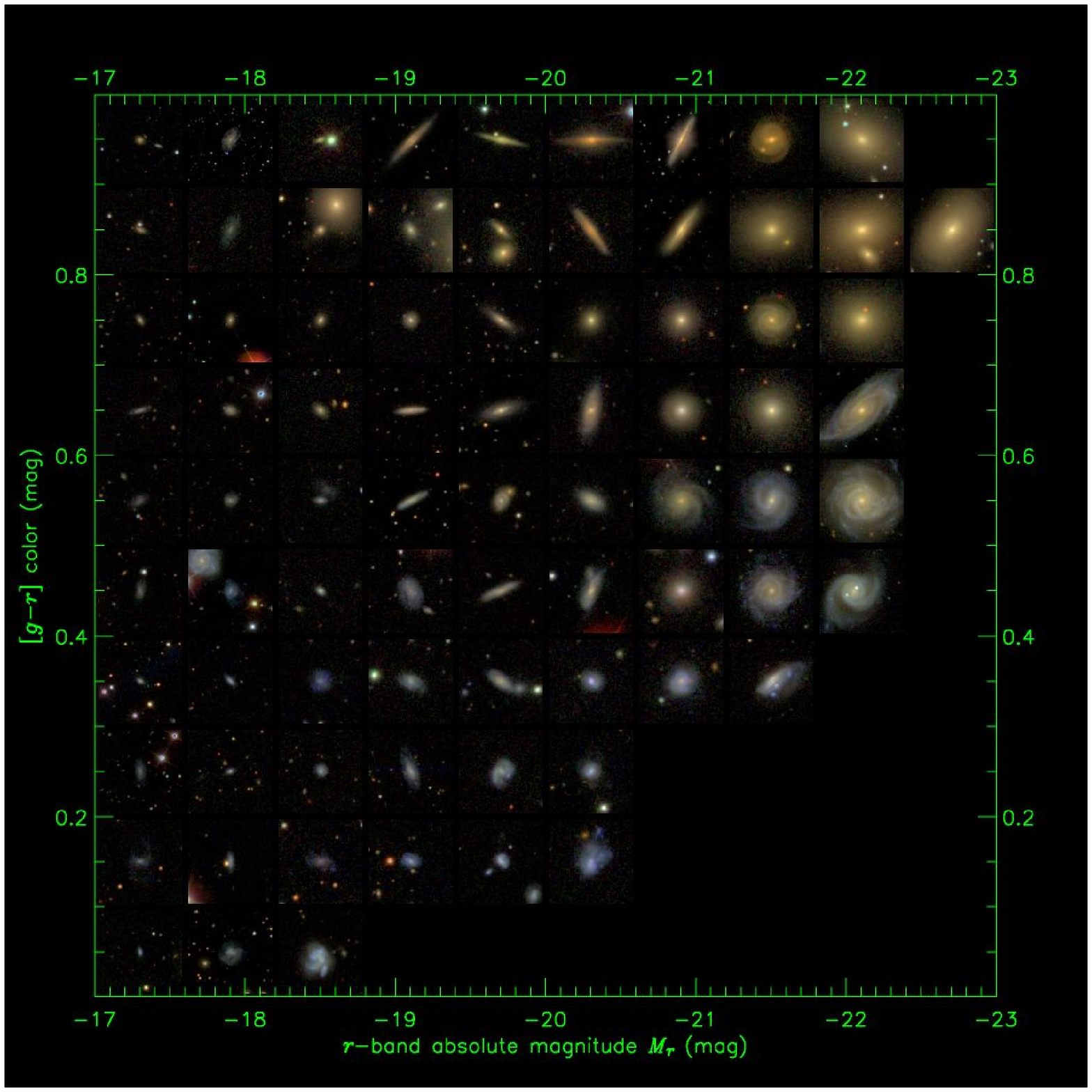}
\caption{\label{color_mag} In each bin of color and absolute
magnitude, we show a randomly chosen SDSS galaxy image. Each image is
40 $h^{-1}$ kpc on a size.}
\end{figure}

We will be studying the environments of galaxies as a function of
their luminosity and color below. To give a sense of the morphological
properties of galaxies with various luminosities and colors, Figure
\ref{color_mag} shows galaxies randomly selected in bins of color and
luminosity. Each image is 40 $h^{-1}$ kpc on a side. Red, high
luminosity galaxies are classic giant ellipticals.  Lower luminosity
red galaxies tend to be more flattened and less concentrated.  Blue,
high luminosity galaxies have well-defined spiral structure and dust
lanes. Lower luminosity blue galaxies have less well-defined bulges
and fewer spiral features.

\subsection{Densities}

In order to evaluate the environments of galaxies in our sample, we
perform the following procedure. 

First, for each given galaxy in the sample, we count the number of
other galaxies $N_i$ with $M_r-5\log_{10} h < -18.5$ outside a
projected radius of 10 $h^{-1}$ kpc and within some outer radius
$r_T$, which we will vary below, and within $\pm 1000$ km s$^{-1}$ in
the redshift direction. This trace catalog is volume-limited within
$z<0.05$. In order to make a more direct comparison to
\citet{balogh04a}, we will also use a trace catalog containing only
galaxies with $M_r-5\log_{10} h  < -20.5$.

Second, we calculate the mean expected number of galaxies in that
volume as:
\begin{equation}
N_{\mathrm{exp}, i} = {\bar{n}} \int\,dV f(\alpha, \delta) 
\end{equation}
where $f(\alpha, \delta)$ is the sampling fraction of galaxies in the
right ascension ($\alpha$) and declination ($\delta$) direction of
each point within the volume. We perform this integral using a Monte
Carlo approach, distributing random points inside the volume with a
density modulated by the sampling fraction $f(\alpha, \delta)$.


In order to calculate the mean density around galaxies in various
classes, we will simply calculate:
\begin{equation}
1+\delta = \frac{\sum_i N_i}{\sum_i N_{\mathrm{exp}, i}}
\end{equation}
as the density with respect to the mean.

\section{Dependence of mean density on color and luminosity}
\label{results}

\begin{figure}
\figurenum{\fignum}
\plotone{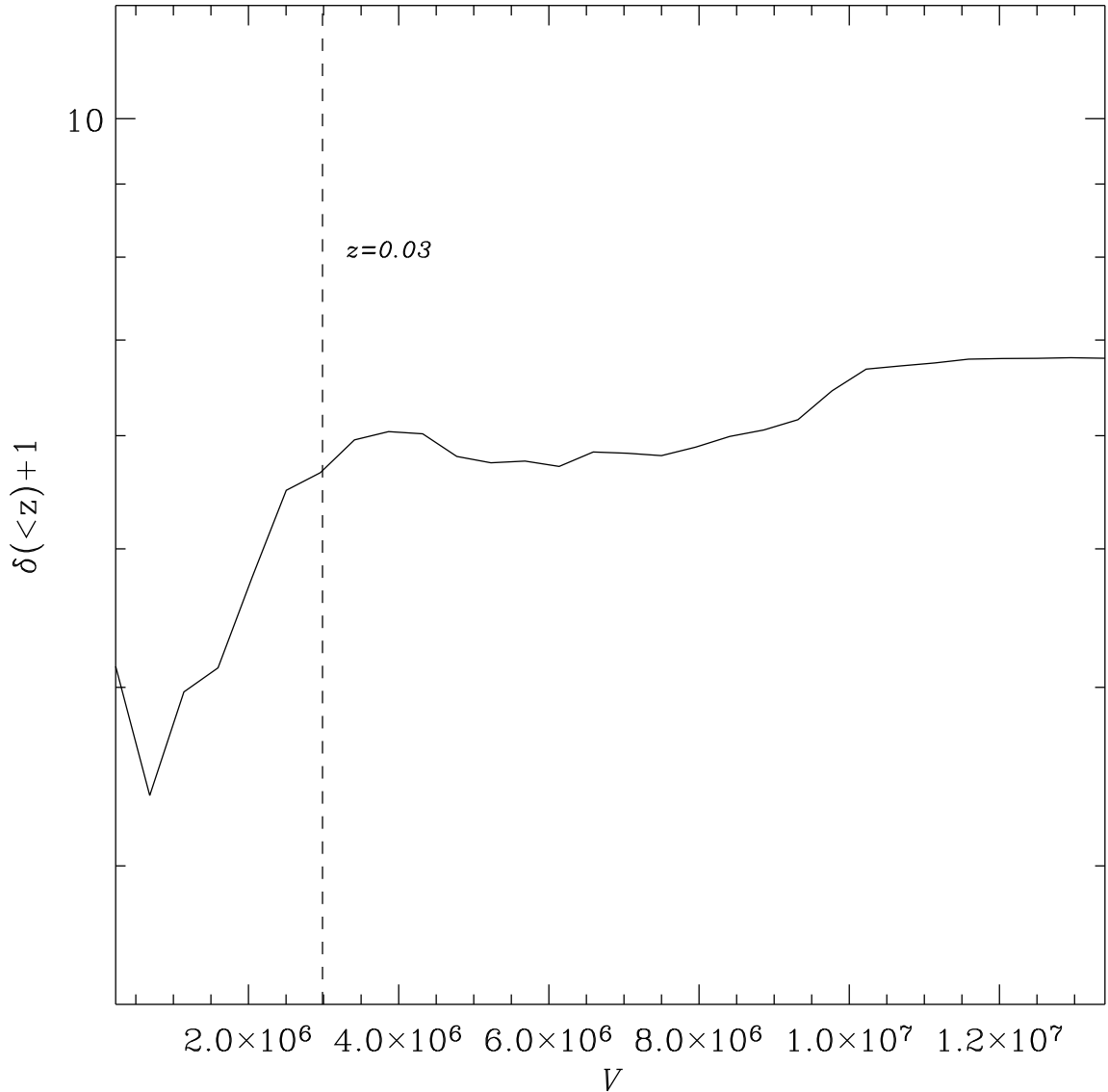}
\caption{\label{check_rho_converge} Mean density $\delta+1$
on scales $r_T = 1.0$ $h^{-1}$ Mpc around galaxies with $M_r < -18.5$
for galaxies at redshifts $<z$, as a function of enclosed volume in
the sample $V(z)$ (in units of $h^{-3}$ Mpc$^3$).  The mean is
calculated in a sliding box with the width shown.  If the sliding box
contains fewer than 20 galaxies, the result is colored pure white.
Such galaxies are observable out to $z=0.05$, the limit of this
plot. The mean density of such galaxies converges at $z=0.03$. }
\end{figure}

When one calculates the mean density around galaxies, it is necessary
to have a fair sample of the universe. For the most luminous galaxies
in our sample ($M_r - 5 \log_{10} h < -18.5$) the sample is
volume-limited out to our redshift limit of $z=0.05$ and constitutes
the equivalent of a 60 $h^{-1}$ Mpc radius sphere, which constitutes a
fair sample for many purposes ($\Lambda$CDM predicts a variance in
such a sphere to be about 0.13). However, the lower luminosity
galaxies can only be seen in the fraction of this volume which is
nearby, and below a certain luminosity the sample is no longer
fair. For example, consider Figure \ref{check_rho_converge}, which
shows the cumulative mean density around galaxies with $M_r- 5
\log_{10} h <-18.5$ in spheres of larger and larger radius around the
Milky Way. The mean overdensity does not converge until a volume which
corresponds to approximately $z=0.03$. Thus, it is not really safe to
evaluate the mean density around galaxies that are too low luminosity
to be observed out that far in redshift, which is to say, less
luminous than $M_r- 5 \log_{10} h = -17$.

\begin{figure}
\figurenum{\fignum}
\plotone{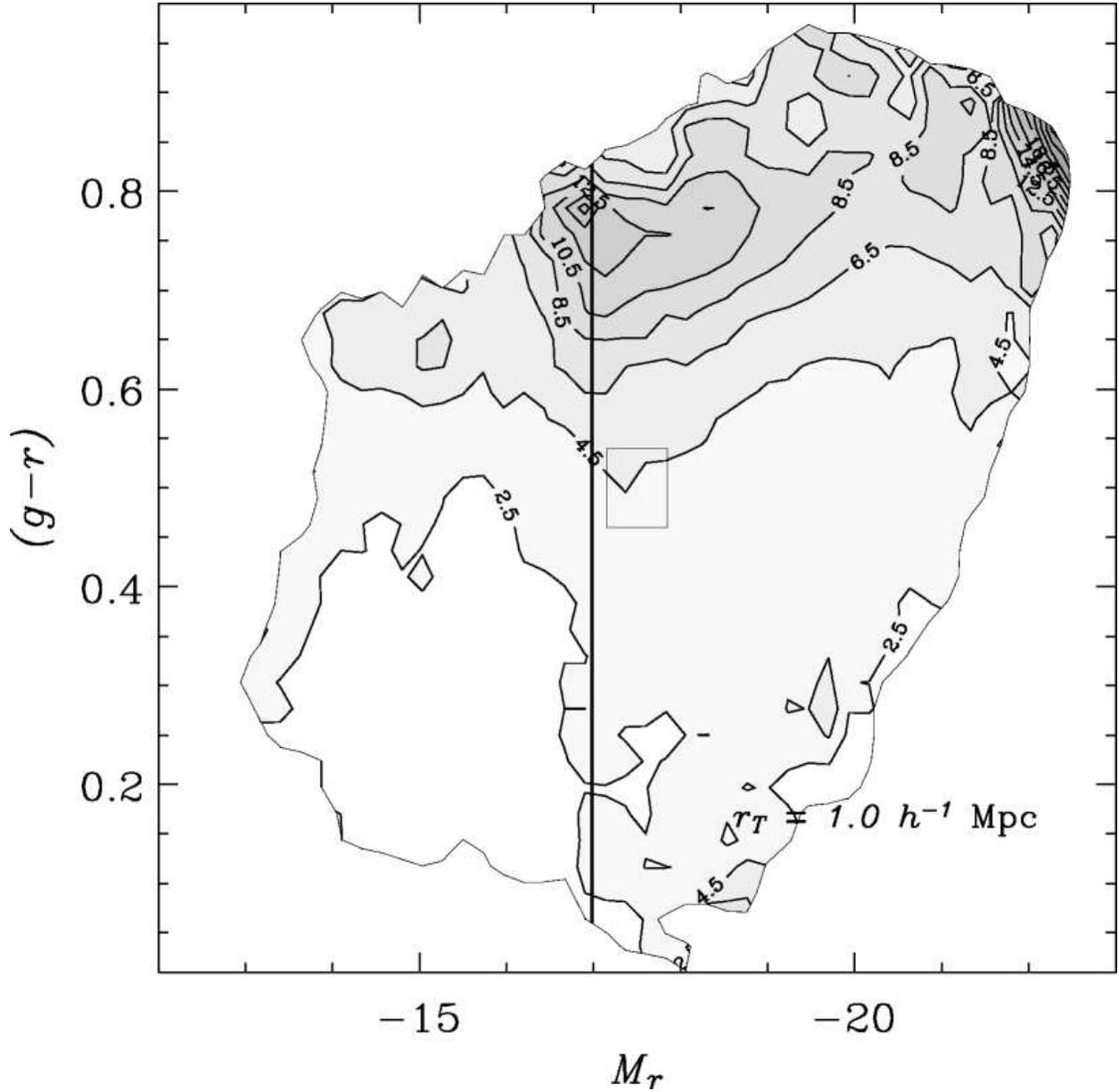}
\caption{\label{biden_all} Mean density $\delta(M_r, g-r)+1$ around
galaxies using $r_T = 1$ $h^{-1}$ Mpc. As found in \citet{hogg03b} and
\citet{blanton03q}, there is a strong increase in the mean density
around luminous galaxies as a function of luminosity, and around blue
galaxies as a function of color. The solid line indicates the minimum
luminosity for which our sample constitutes a fair sample (to the left
of this line the mean density of galaxies is underestimated).}
\end{figure}

However, for the moment let us consider Figure \ref{biden_all}. The
greyscale and contours show the mean density relative to the mean as a
function of color and luminosity, using a projected radius of $0.5$
$h^{-1}$ Mpc. The mean is calculated in a sliding box with the width
shown.  If the sliding box contains fewer than 20 galaxies, the result
is ignored and colored pure white. Here we show the results for the
entire sample. Our statistical uncertainties are well-behaved down to
about $M_r- 5 \log_{10} h \sim -14.5$, but we are likely to be cosmic
variance limited for $M_r- 5 \log_{10} h > -17$, as indicated by the
solid vertical line. Thus, the apparent decline in the mean
overdensity for red galaxies lower luminosity than $M_r- 5 \log_{10} h
\sim -17$ is probably spurious. Despite that limitation, we note 
that there is a strong relationship between environment and color even
at $M_r- 5 \log_{10} h \sim -15$.

We note in passing that we can still use the variation of the density
within $z<0.03$ to study the properties of galaxies as a function of
density down to low luminosity. Just because the {\it mean} density of
galaxies in that volume has not converged does {\it not} imply that
there is insufficient variation of density to study the variation of
galaxy properties with environment.

\begin{figure}
\figurenum{\fignum}
\plotone{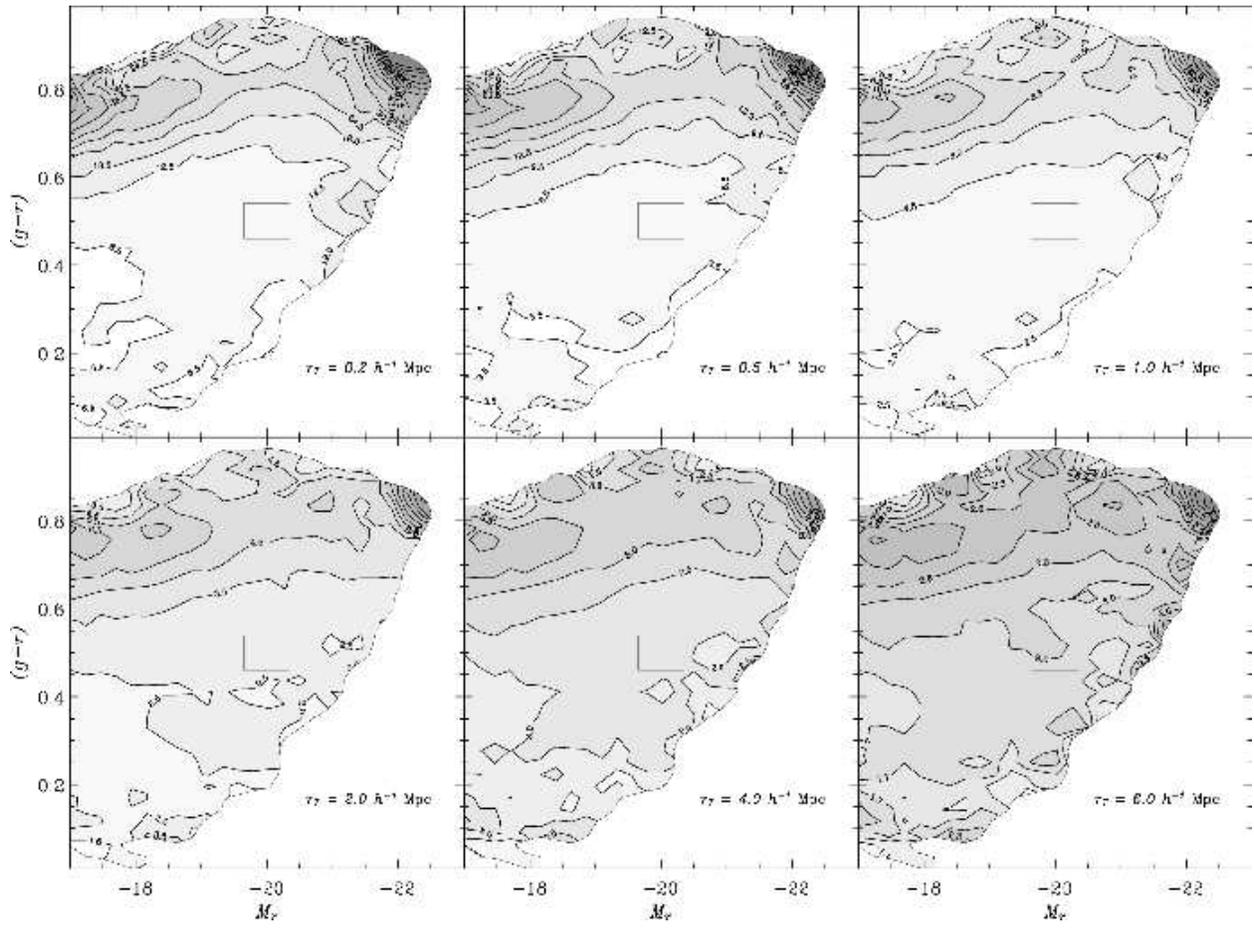}
\caption{\label{biden_scales} Similar to Figure \ref{biden_all}, now
limited to $M_r < -17$, and showing six different scales.}
\end{figure}

For our fair sample of galaxies with $M_r- 5 \log_{10} h <-17$, Figure
\ref{biden_scales} shows the dependence of overdensity on luminosity
and color for six different projected radii: 0.2, 0.5, 1, 2, 4, and 6
$h^{-1}$ Mpc. We only show results for $M_r- 5 \log_{10} h < -17$, for
which we have a fair sample. Obviously, the density contrast decreases
with scale; on the other hand, the qualitative form of the plot does
not change.

Our results remain similar to those shown in \citet{hogg03b} and
\citet{blanton03q}. The results here demonstrate that the environments
of low luminosity, red galaxies do continue to become denser as
absolute magnitude increases down to absolute magnitudes of $-17$
(about two magnitudes less luminous than explored by our previous
work).

\begin{figure}
\figurenum{\fignum}
\plotone{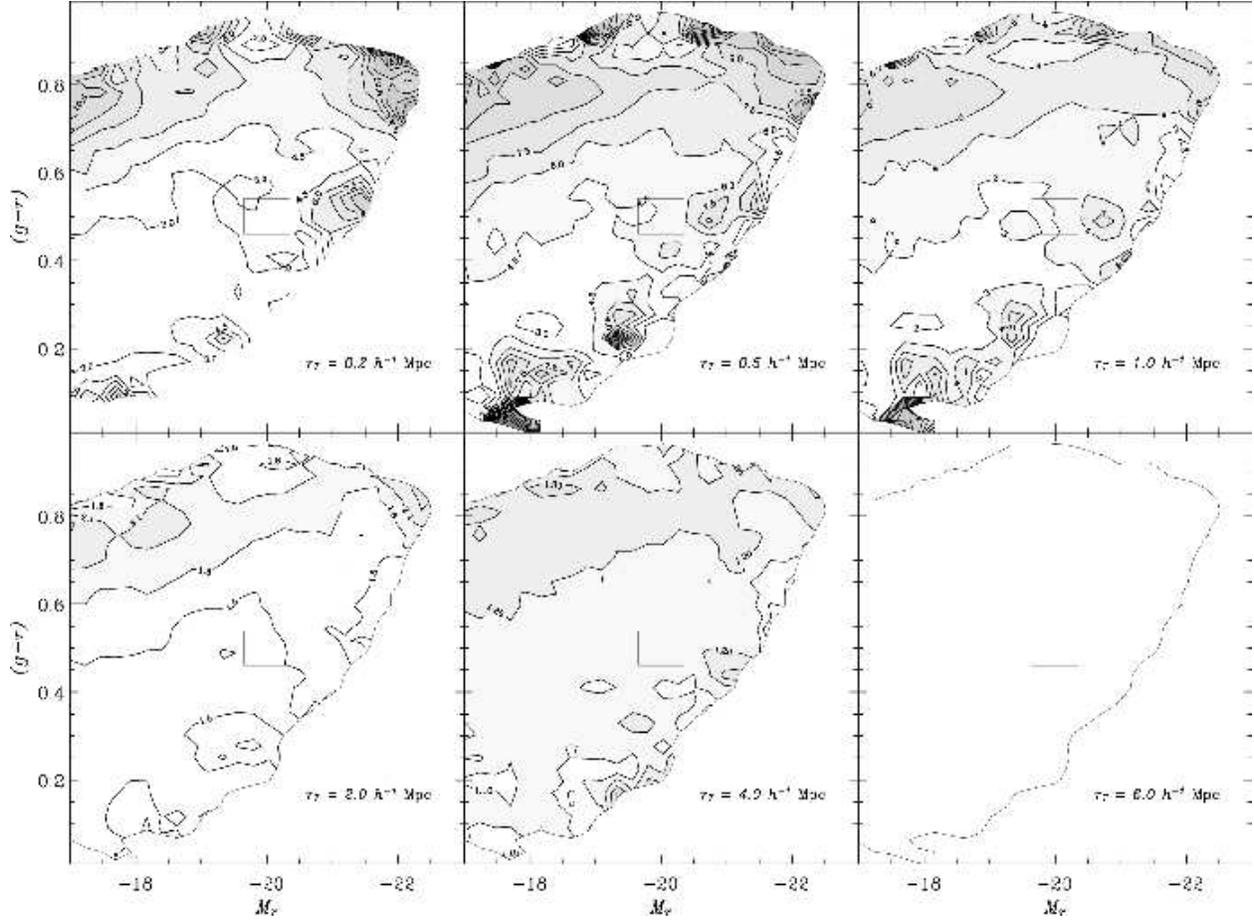}
\caption{\label{biden_ratios} Similar to Figure \ref{biden_scales},
but showing the ratio of the overdensity $\delta(M_r, g-r)$ at each
scale to that at $r_T = 6$ $h^{-1}$ Mpc. Thus, the bottom right panel
is unity across the entire plot. This ratio is a measure of the
steepness of the correlation function at each absolute magnitude and
color. Interestingly, the contours of steepness are similar to the
contours in overdensity at any scale, meaning that wherever the local
density at $r_T=6.0$ $h^{-1}$ Mpc is higher, the correlation function
is also steeper. }
\end{figure}

Figure \ref{biden_ratios} shows the ratio of the overdensity $\delta$
at each scale relative to that at the largest scale of $r_T = 6$
$h^{-1}$ Mpc. This ratio is a measure of the steepness of the
cross-correlation between galaxies of a given color and absolute
magnitude with {\it all} galaxies in our volume-limited sample ($M_r-
5 \log_{10} h < -18.5$). Interestingly, the contours in steepness are
qualitatively similar to the contours in overdensity in Figure
\ref{biden_scales}. This similarity implies that for each class of 
galaxy, the strength of the correlation on large scales always is
associated with a {\it steeper} correlation function.

\section{Blue fraction as a function of environment}
\label{bluefrac}

\begin{figure}
\figurenum{\fignum}
\plotone{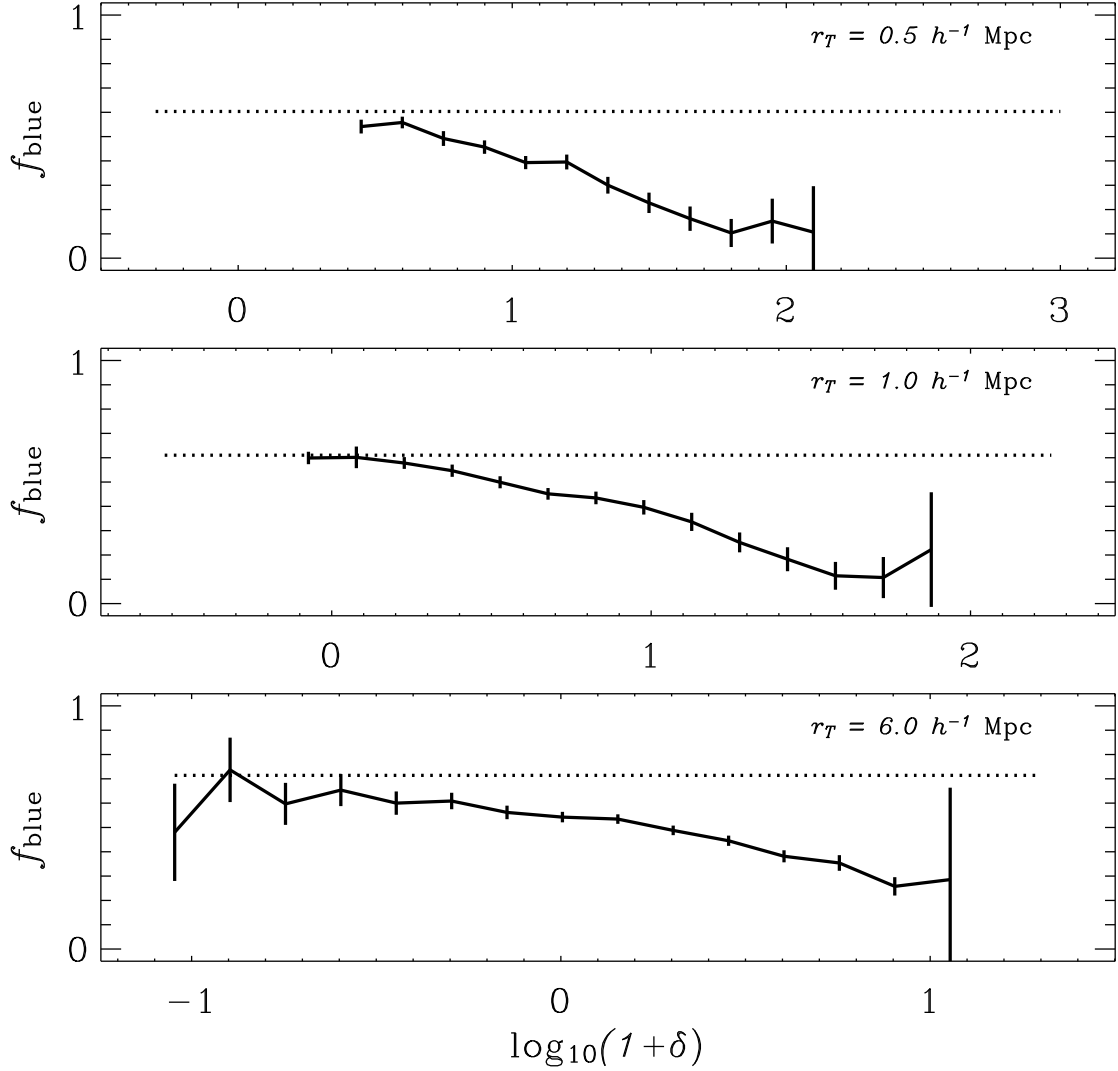}
\caption{\label{lowz_fracblue} Fraction of blue galaxies (using the
definition of Equation \ref{colorcut}) as a function of local density
on three different scales, as labeled. Uncertainties are Poisson
estimates (not binomial). The dotted line represents the blue fraction
for all galaxies with no neighbors. }
\end{figure}

Another way of looking at similar results is to ask, as a function of
environment, what fraction of galaxies are blue. We split the sample
into ``red'' and ``blue'' galaxies using the following,
luminosity-dependent cut:
\begin{equation}
\label{colorcut}
(g-r)_c = 0.65 - 0.03 (M_r-5\log_{10} h +20)
\end{equation}
Blue galaxies thus have $(g-r) < (g-r)_c$. We then sort all the
galaxies with $M_r- 5 \log_{10} h < -18.5$ into bins of density on
three different scales: $r_T = 0.5$, $1.0$, and $6.0$ $h^{-1}$ Mpc.
In each bin we calculate the fraction of blue galaxies.  Figure
\ref{lowz_fracblue} shows this blue fraction as a function of density.
In all cases, the blue fraction declines as a function of density, as
one would expect based on Figure \ref{biden_scales} above, and from
the astronomical literature (a highly abridged list of relevant work
would include \citealt{hubble36a, oemler74a, dressler80a, hermit96a,
guzzo97a, giuricin01a, hashimoto99a, norberg02a, zehavi04a}). If we
divide the sample into bins of luminosity, we find that higher
luminosities have smaller blue fractions (of course) but that the
dependence of blue fraction on density does not change.

\begin{figure}
\figurenum{\fignum}
\plotone{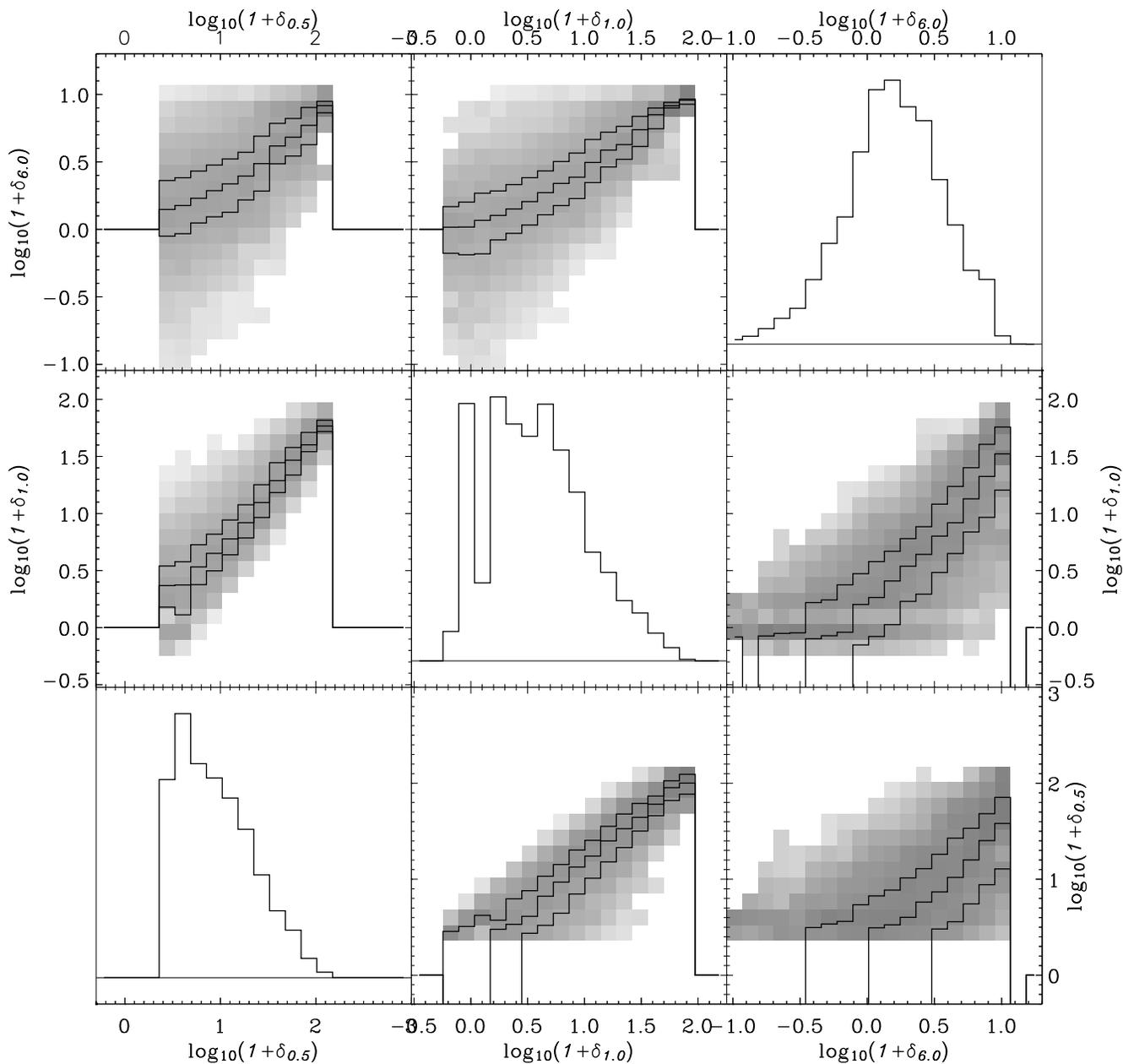}
\caption{\label{denvden} Comparison of density estimates on different
scales. Diagonal plots show the distribution of density estimates on
scales of $r_T = 0.5$, $1.0$, and $6.0$ $h^{-1}$
Mpc. Off-diagonal plots show as a greyscale the conditional
distribution of each density estimate with respect to all the
others. The lines show the quartiles of the distribution. Generally
speaking, a larger density on one scale indicates a larger density on
all other scales. Obviously, the correlation is stronger for scales
closer to one another. Note that while a large density for $r_T =
0.5$ $h^{-1}$ Mpc implies a large density for $r_T = 6.0$ $h^{-1}$
Mpc, the opposite is not the case: there is a large range of densities
on the smaller scale given {\it any} value of the density on the larger
scale. }
\end{figure}

The question naturally arises: {\it which} scales are important to the
process of galaxy formation? Is the local environment within 0.5
$h^{-1}$ Mpc the only important consideration? Or is the larger scale
environment also important? For example, consider Figure
\ref{denvden}, which shows the conditional dependence of the three
density estimators at the three scales on each other. The diagonal
plots simply show the distribution within our sample of each density
estimator. The off-diagonal plots show the conditional distribution of
the quantity on the $y$-axis given the quantity on the $x$-axis. As an
example, the lower right panel shows $P\left(\log_{10}(1+\delta_{0.5})
| \log_{10}(1+\delta_{6.0})\right)$. The lines are the quartiles of
the distribution. Obviously, the estimators are correlated with one
another; thus, a dependence of blue fraction on one on them is likely
to cause a dependence of blue fraction on any of the density
estimators. However, physically, our theoretical expectation is that
the density on smaller scales is more important than that on larger
scales. That is, for a given density on small scales, we expect that
the density on much larger scales (much larger than the size of the
largest virialized dark matter halos) should not be closely related to
the properties of the galaxies.  To address this question, we ask what
the blue fraction is as a function of density measurements on two
different scales.

\begin{figure}
\figurenum{\fignum}
\plotone{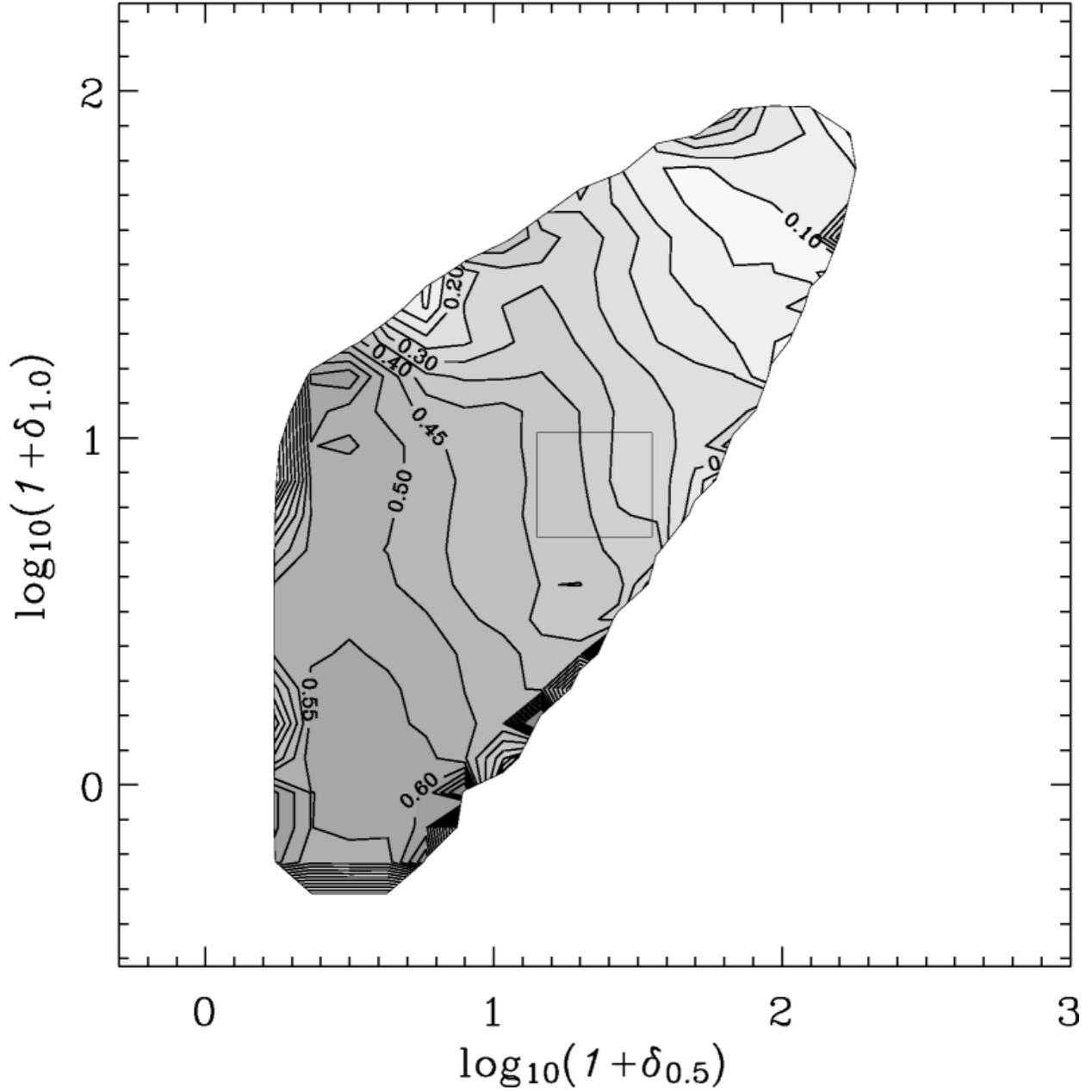}
\caption{\label{lowz_fracblue2_0.5-1.0} Contours and greyscale show
 the blue galaxy fraction as a function of local density on $r_T =
 0.5$ and $1$ $h^{-1}$ Mpc scales. The local density on both scales is
 important. }
\end{figure}

Figure \ref{lowz_fracblue2_0.5-1.0} shows the fraction of blue
galaxies as a function of two density estimates: one with $r_T = 0.5$
$h^{-1}$ Mpc and one with $r_T = 1.0$ $h^{-1}$ Mpc . In this case it
is clear that the blue fraction is a function of both densities. That
is, even at a fixed density on scales of $0.5$ $h^{-1}$ Mpc, the
density outside that radius matters to the blue fraction; in addition,
at a fixed density on scales of $1.0$ $h^{-1}$ Mpc, the distribution
of galaxies within that radius appears to affect the blue fraction as
well.

\begin{figure}
\figurenum{\fignum}
\plotone{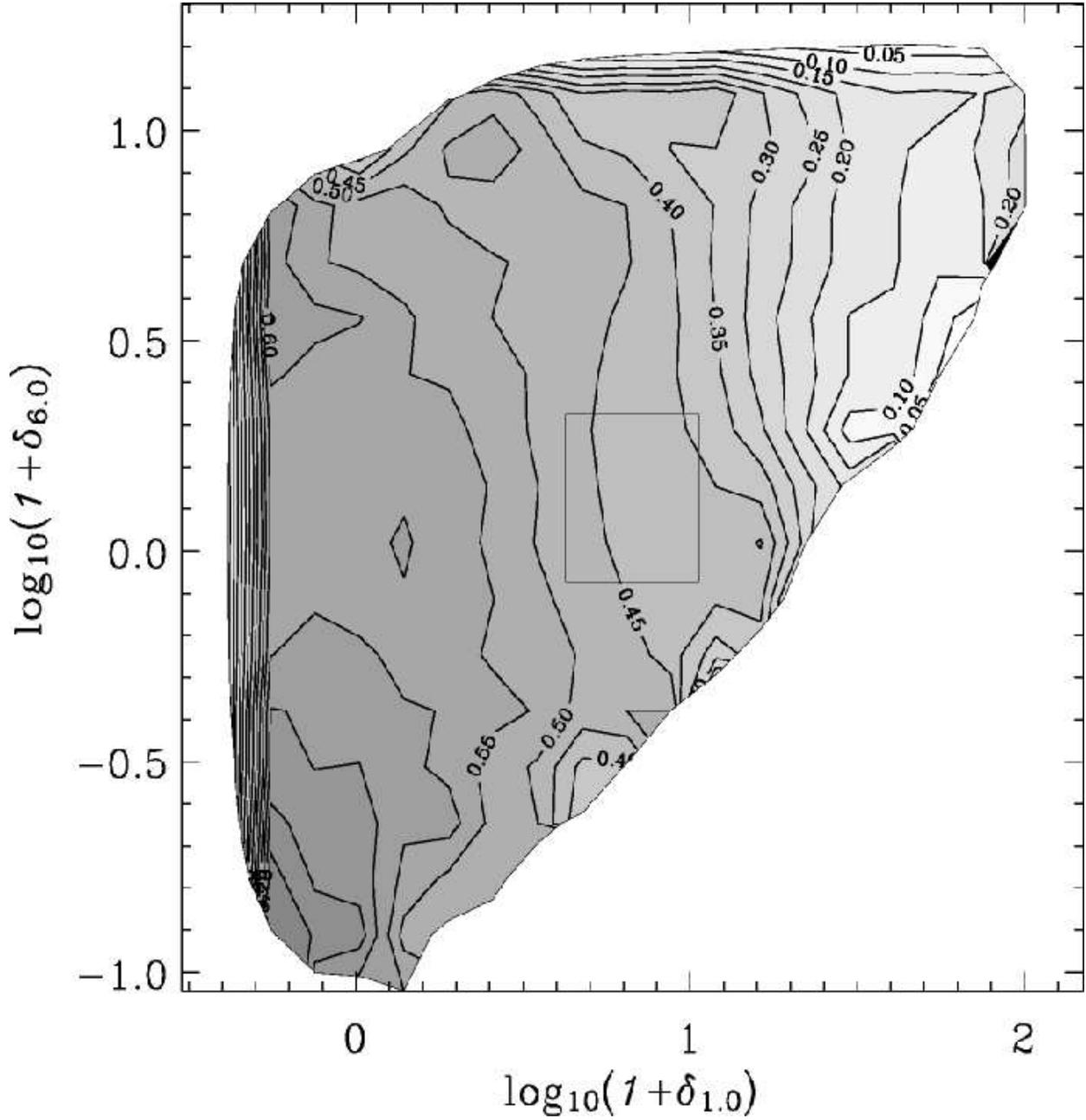}
\caption{\label{lowz_fracblue2_1.0-6.0} Contours and greyscale show
 the blue galaxy fraction as a function of local density on $r_T = 1$
 and $6$ $h^{-1}$ Mpc scales. The vertical contours demonstrate that
 only the local density on the smaller scale is important. }
\end{figure}

On the other hand, consider Figure \ref{lowz_fracblue2_1.0-6.0}, which
is the same as Figure \ref{lowz_fracblue2_0.5-1.0}, but now showing
the densities at scales of $r_T = 1.0$ and $6.0$ $h^{-1}$ Mpc. In
Figure \ref{lowz_fracblue2_1.0-6.0} the contours are vertical,
indicating that the density between $1.0$ and $6.0$ $h^{-1}$ Mpc has
very little effect on galaxy properties. At a fixed value of the
density at the smaller scale, the larger scale environment appears to
be of little importance.

\begin{figure}
\figurenum{\fignum}
\plotone{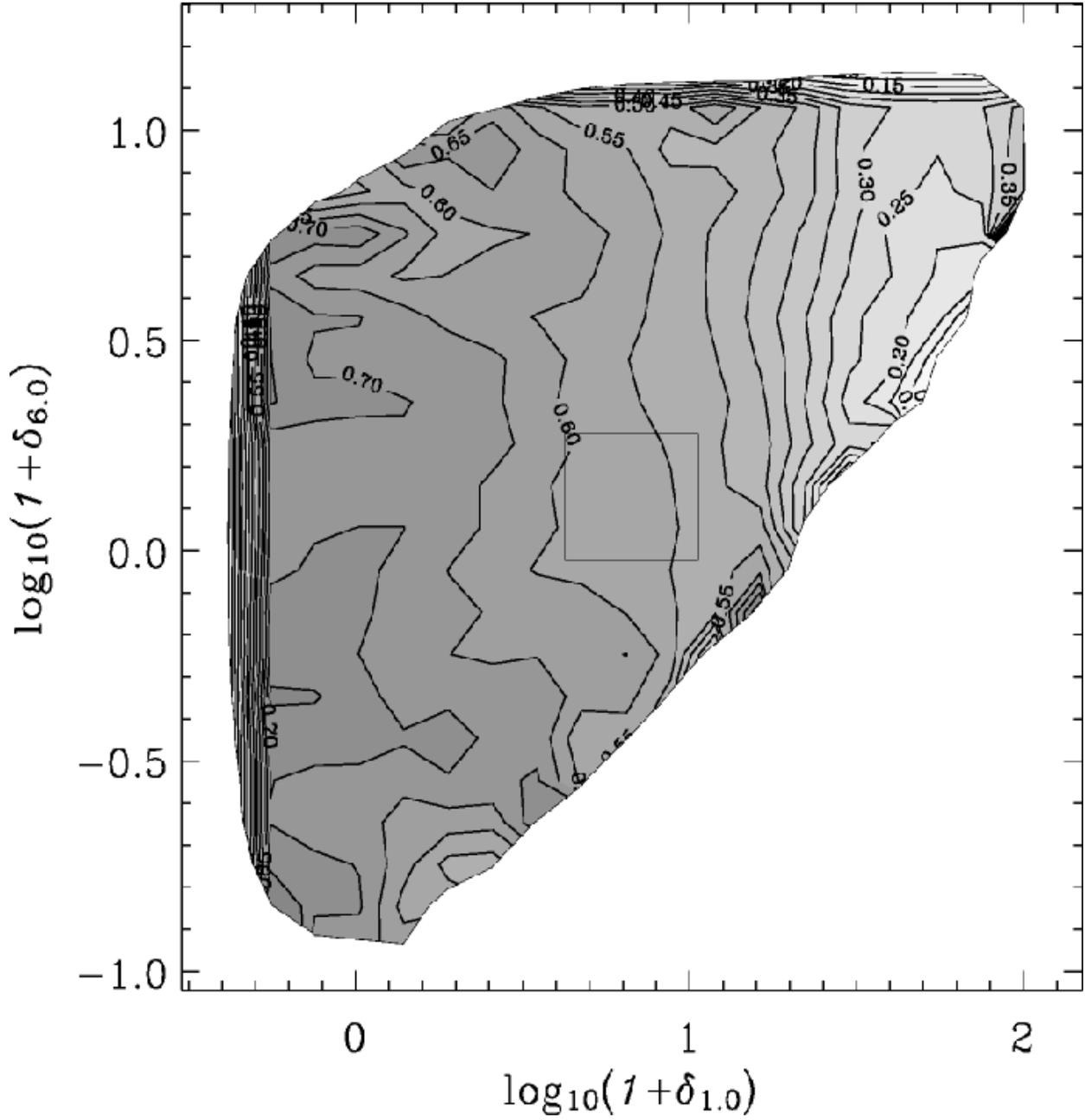}
\caption{\label{lowz_frachalpha2_1.0-6.0} Similar to Figure
\ref{lowz_fracblue2_1.0-6.0}, but for the H$\alpha$ emitting fraction
rather than the blue fraction. Again, only the small scales are
important.} 
\end{figure}

\citet{balogh04a} found that these contours were not vertical when he
looked at the fraction of galaxies for which the H$\alpha$ equivalent
width was $> 4$ \AA. Their result appears in conflict with that of the
previous paragraph.  On the other hand, the emission lines measure a
more recent star formation rate than does the color; it is possible in
principle that the more recent star formation rate depends more
strongly on large-scale environment. To rule out this possibility,
Figure \ref{lowz_frachalpha2_1.0-6.0} shows the same result as Figure
\ref{lowz_fracblue2_1.0-6.0}, but now showing the fraction of galaxies
with H$\alpha$ equivalent widths (as measured by
\citealt{tremonti04a}) greater than 4 \AA. Again, for strong emission
line fraction as for the blue fraction, the smaller scales are
important, but the 6 $h^{-1}$ Mpc scales are not, in contradiction
with \citet{balogh04a}. 

\begin{figure}
\figurenum{\fignum}
\plotone{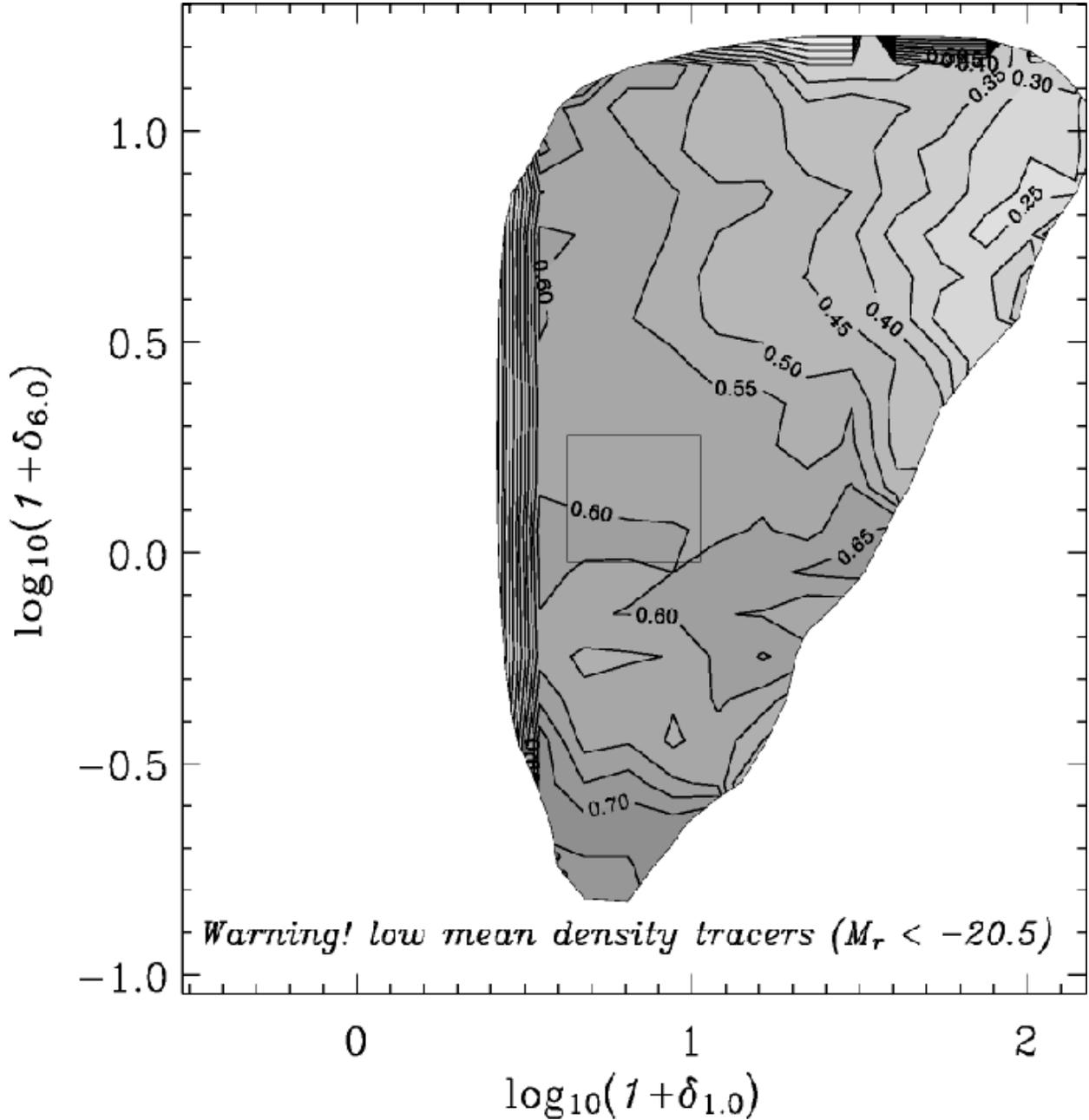}
\caption{\label{lowz_frachalpha2_M-20.5_1.0-6.0} Similar to Figure
\ref{lowz_frachalpha2_1.0-6.0}, but (for comparison with
\citealt{balogh04a}) using galaxies with $M_r < -20.5$
as tracers rather than galaxies with $M_r < -18.5$. With these
tracers, the large scale density field yields extra information not
contained in the small scale density field, probably because there is
effectively ``noise'' in the relationship between the brighter, lower
mean density tracers and the underlying density field which determines
the properties of the galaxy population.}
\end{figure}

Why, then, did \citet{balogh04a} conclude that large scales {\it were}
important? There are a number of differences between our study and
theirs. First, their contouring method differs; instead of measuring
the blue fraction in bins of fixed size, at each point they measure
the star-forming fraction among the nearest 500 galaxies in the plane
of $\delta_{1.1}$ and $\delta_{5.5}$. We have found that this
procedure creates a {\it slight} bias in the contouring in the sense
that near the edges of the distribution vertical contours will become
diagonal. However, this effect is not strong enough to explain the
differences between our results and those of
\citet{balogh04a}. Second, to estimate the density in their sample
they used a spherical Gaussian filter, whereas here we use the
overdensity in cones. We have not investigated what effect this
difference has. Finally, they use tracer galaxies with a considerably
lower mean density than ours. Their effective absolute magnitude limit
is $M_r-5\log_{10} h < -20.5$; such galaxies have a mean density of
$\sim 4.0 \times 10^{-3}$ $h^{3}$ Mpc$^{-3}$. Our tracers
($M_r-5\log_{10} h < -18.5$) have a mean density of $\sim 2.3 \times
10^{-2}$ $h^{3}$ Mpc$^{-3}$, almost six times higher.  Figure
\ref{lowz_frachalpha2_M-20.5_1.0-6.0} shows our results when we
restrict our tracer sample to $M_r-5\log_{10} h < -20.5$.  The
contours in this figure are very diagonal, similar to the results of
\citet{balogh04a}. 

This result suggests that one of two possible mechanisms are causing
the differences between our results and those of
\citet{balogh04a}. First, the higher luminosity galaxies with
$M_r-5\log_{10} h < -20.5$ might be yielding fundamentally different
information about the density field than our lower luminosity
tracers. Second, the lower mean density of the galaxies with
$M_r-5\log_{10} h < -20.5$ might be effectively introducing ``noise''
in the measurement on small scales. Remember that the large scale and
small scale densities are intrinsically correlated. So if the small
scale measurement is noisy enough, the higher signal-to-noise ratio
large scale measurement could actually be adding extra information
about the environment on {\it small} scales. Such an effect would make
the contours in Figure \ref{lowz_frachalpha2_M-20.5_1.0-6.0} diagonal.
We have performed a simple test to distinguish these possibilities,
which is to remake Figure \ref{lowz_frachalpha2_1.0-6.0} using the low
luminosity tracers ($M_r -5\log_{10} h < -18.5$) but subsampling them
to the same mean density as the high luminosity tracers ($M_r
-5\log_{10} h < -20.5$). This test yields diagonal contours, meaning
one can understand the diagonal contours of Figure
\ref{lowz_frachalpha2_M-20.5_1.0-6.0} and of \citet{balogh04a} as
simply reflecting the low signal-to-noise ratio of the density
estimates.

\section{Summary and Discussion}
\label{summary}

We explore the relative bias between galaxies as a function of scale,
finding the following.
\begin{enumerate}
\item The dependence of mean environment on color persists to the
lowest luminosities we explore ($M_r-5\log_{10} h  \sim -14$).
\item Red, low luminosity galaxies tend to be in overdense regions, down
at least to $M_r-5\log_{10} h  \sim -17$. This result extends those found by
\citet{hogg03b} and \citet{blanton03q} towards lower luminosities by
about 2 magnitudes.
\item At any given point of color and luminosity, a 
correlation function with a stronger amplitude implies correlation
function with a steeper slope.
\item In regions of a given overdensity on small scales ($r_T = 1$
$h^{-1}$ Mpc), the overdensity on large scales ($r_T = 6$ $h^{-1}$
Mpc) does not appear to relate to the recent star formation history of
the galaxies. 
\end{enumerate}

The last point above deserves elaboration. First, it contradicts the
results of \citet{balogh04a}. We have found that their results are
probably due to the low mean density of the tracers they used. This
explanation underscores the importance of taking care when using low
signal-to-noise quantities. Galaxy environments are difficult to
measure, in the sense we use tracers that do not necessarily trace the
``environment'' perfectly, meaning neither with low noise nor
necessarily in an unbiased manner. We claim here that our higher
density of tracers marks an improvement over previous work, but it is
worth noting the limitations of assuming that the local galaxy density
fairly and adequately represents whatever elements of the environment
affect galaxy formation.

Second, if the galaxy density field is an adequate representation of
the environment, the result has important implications regarding the
physics of galaxy formation.  In simulations whose initial conditions
are constrained by cosmic microwave background observations and galaxy
large-scale structure observations, virialized dark matter halos do
not extend to sizes much larger than $1$--$2$ $h^{-1}$ Mpc. Thus, our
results are consistent with the notion that only the masses of the
host halos of the galaxies we observe are strongly affecting the star
formation of the galaxies. In addition,
\citet{blanton03q} find that only the star formation histories, not
the azimuthally-averaged structural parameters, are directly related
to environment. For these reasons, it is likely that we can understand
the process of galaxy formation by only considering the properties of
the host dark matter halos. Our results therefore encourage the ``halo
model'' description of galaxy formation and the pursuit of
semi-analytic models which depend only on the properties of the host
halo ({\it e.g.}, \citealt{kauffmann97a, seljak00, benson01a, white01a,
berlind02a}).

\acknowledgments

Thanks to Eric Bell and George Lake for useful discussions during
this work. Thanks to Guinevere Kauffmann for encouraging us to pursue
this question. Thanks to Christy Tremonti and Jarle Brinchmann for
the public distribution of their measurements of SDSS spectra.

Funding for the creation and distribution of the SDSS has been
provided by the Alfred P. Sloan Foundation, the Participating
Institutions, the National Aeronautics and Space Administration, the
National Science Foundation, the U.S. Department of Energy, the
Japanese Monbukagakusho, and the Max Planck Society. The SDSS Web site
is {\tt http://www.sdss.org/}.

The SDSS is managed by the Astrophysical Research Consortium (ARC) for
the Participating Institutions. The Participating Institutions are The
University of Chicago, Fermilab, the Institute for Advanced Study, the
Japan Participation Group, The Johns Hopkins University, Los Alamos
National Laboratory, the Max-Planck-Institute for Astronomy (MPIA),
the Max-Planck-Institute for Astrophysics (MPA), New Mexico State
University, University of Pittsburgh, Princeton University, the United
States Naval Observatory, and the University of Washington.

\newpage

\clearpage
\clearpage

\setcounter{thefigs}{0}

\clearpage
\stepcounter{thefigs}

\clearpage
\stepcounter{thefigs}

\clearpage
\stepcounter{thefigs}

\clearpage
\stepcounter{thefigs}

\clearpage
\stepcounter{thefigs}

\clearpage
\stepcounter{thefigs}

\clearpage
\stepcounter{thefigs}

\clearpage
\stepcounter{thefigs}

\clearpage
\stepcounter{thefigs}

\clearpage
\stepcounter{thefigs}

\clearpage
\stepcounter{thefigs}

\newpage

\end{document}